# Exponential Random Graph Modeling for Complex Brain Networks

**Sean L. Simpson[1*], Satoru Hayasaka[1,2], Paul J. Laurienti[2]**

**1** Department of Biostatistical Sciences, Wake Forest University School of Medicine, Winston-Salem, NC, USA,

**2** Department of Radiology, Wake Forest University School of Medicine,Winston-Salem, NC, USA

[*]E-mail: slsimpso@wfubmc.edu

# Abstract

Exponential random graph models (ERGMs), also known as p* models, have been utilized extensively in the social science literature to study complex networks and how their global structure depends on underlying structural components. However, the literature on their use in biological networks (especially brain networks) has remained sparse. Descriptive models based on a specific feature of the graph (clustering coefficient, degree distribution, etc.) have dominated connectivity research in neuroscience. Corresponding generative models have been developed to reproduce one of these features. However, the complexity inherent in whole-brain network data necessitates the development and use of tools that allow the systematic exploration of several features simultaneously and how they interact to form the global network architecture. ERGMs provide a statistically principled approach to the assessment of how a set of interacting local brain network features gives rise to the global structure. We illustrate the utility of ERGMs for modeling, analyzing, and simulating complex whole-brain networks with network data from normal subjects. We also provide a foundation for the selection of important local features through the implementation and assessment of three selection approaches: a traditional p-value based backward selection approach, an information criterion approach (AIC), and a graphical goodness of fit (GOF) approach. The graphical GOF approach serves as the best method given the scientific interest in being able to capture and reproduce the structure of fitted brain networks.

## Introduction

### Brain networks

Whole-brain connectivity analyses are gaining prominence in the neuroscientific literature due to the need to understand how various regions of the brain interact with one another. The inherent complexity in the way these regions interact necessitates studying the brain as a whole rather than just its individual parts. The application of network and graph theory to the brain has facilitated these whole-brain analyses and helped to uncover new insights into the structure and function of the nervous system. Structural and functional connectivity studies have revealed that the brain exhibits the small-world properties [1-4]. These properties are characterized by tight local clustering and efficient long distance connections as described in the seminal work of [5]. Network models based on a given small-world property or other local property (e.g., node degree ($k$)) have mostly been utilized as a means to describe various brain networks. However, in order to gain deeper insights into the complex neurobiological interactions and changes that occur in many neurological conditions and disorders, analysis methods that enable systematically assessing several properties simultaneously are needed given the statistical dependencies among these properties [6,7]. The exponential random graph models discussed in this paper provide one such analysis approach.

### Exponential random graph models

Exponential random graph models (ERGMs), also known as p* models [8-11], have been utilized extensively in the social science literature to analyze complex network data as discussed in [12,13] and others. However, the literature on their use in biological networks (especially brain networks) has remained sparse. Descriptive models based on a specific feature of the network such as characteristic path length ($L$) and clustering coefficient ($C$) have dominated connectivity research in neuroscience [14]. The few inferential studies have employed relatively rudimentary testing techniques such as the ANOVA used in [15] to examine group differences based on one of these features. ERGMs provide a statistically principled approach to the systematic exploration of several features simultaneously and how they interact to form the global network architecture. They allow parsimoniously modeling the probability mass function (pmf) for a given class of graphs based on a set of explanatory metrics (local features). The pmf can then be used to determine the probability that any given graph is drawn from the same distribution as the observed graph. These models enable achieving an efficient representation of complex network data structures and allow examining the way in which a network's global structure and function depend on its local structure. That is, they provide a means of assessing how and to what extent combinations of local (brain) structures produce global network properties.

In ERGMs networks are analogous to a multivariate response variable in regression analysis, with the explanatory metrics quantifying local features of the network such as how clustered connections are (short distance communication) or how well the network transmits information globally (long distance communication). Fitted parameter values from the model can then be utilized to understand particular emergent behaviors of the network (how local features give rise to the global structure). These values can also be used to simulate random realizations of networks that retain constitutive characteristics of the original network.

A more intuitive way to view ERGMs in the brain network context are as models that quantify the relative significance of various graph/network measures ($k$, $C$, $L$, etc.), or their analogues, in explaining the overall network structure, thus enabling generative conjectures about global architecture. These models provide several benefits for brain network researchers. They allow asking specific questions about processes that may give rise to the network architecture via the inclusion of explanatory metrics of choice. ERGMs inherently account for any confounding bias, like the ($N$,$k$)-dependence of network measures (where $N$ is the number of nodes and $k$ the average degree) detailed in [6], when the potential confounding variables are included in the model. The stochastic nature of the model allows understanding and quantifying the uncertainty (an intrinsic feature of complex biological processes) associated with our observed brain network(s) [12]. Simulations based on ERGM fits to brain networks (sets of selected network measures and their parameter estimates) can provide insight into biological variability via the distribution of possible brain networks produced. However, currently, the computational intensiveness of fitting ERGMs may preclude their use with very large networks (e.g., voxel-based networks with tens of thousands of nodes) and certain combinations of network measures.

Here we illustrate the utility of ERGMs for modeling, analyzing, and simulating complex whole-brain network. We also provide a foundation for the development of a "best assessment" ERGM for analyzing complex brain networks. Appropriate statistical comparisons between networks (or groups of networks) via ERGMs necessitates establishing one model (set of explanatory metrics/local features) in order to extract comparable parameter estimates due to the dependence of these features on each other. Toward this end, we assess three potential methods of feature selection for ERGMs in the brain network context. These approaches include a traditional p-value based backward selection approach, an information criterion approach (AIC), and a graphical goodness of fit (GOF) approach. Although the latter two techniques have been discussed in the context of ERGMS [16,17], no detailed comparisons have been performed to determine whether the approaches generally produce the same "best" model/set of features.

## Materials and Methods

### Ethics statement

This study included 10 volunteers representing a subset of a previous study [18]. The study protocol, including all analyses performed here, was approved by the Wake Forest University School of Medicine Institutional Review Board. All subjects gave written informed consent in accordance with the Declaration of Helsinki.

### Data and network construction

Our data include whole-brain functional connectivity networks for 10 normal subjects aged 20-35 (5 female, average age 27.7 years old [4.7 SD]). Each network is comprised of 90 nodes corresponding to the 90 brain regions (90 ROIs-Regions of Interest) defined by the Automated Anatomical Labeling atlas (AAL; [19]). The whole-brain networks were constructed based on fMRI images using graph theory methods. For each subject, 120 images were acquired during 5 minutes of resting using a gradient echo echoplanar imaging (EPI) protocol with TR/TE=2500/40 ms on a 1.5 T GE twin-speed LX scanner with a birdcage head coil (GE Medical Systems, Milwaukee, WI). The acquired images were motion corrected, spatially normalized to the MNI (Montreal Neurological Institute) space and re-sliced to 4×4×5 mm voxel size using an in-house processing script based on SPM99 package (Wellcome Trust Centre for Neuroimaging, London, UK). The resulting images were not smoothed in order to avoid artificially introducing local spatial correlation [20].

The first step in performing the network construction was to generate a whole brain connectivity matrix, or adjacency matrix $(A_{ij})$. This is a binary $n \times n$ matrix where $n$ is the number of nodes representing 90 ROIs. The matrix notes the presence or absence of a connection between any two nodes ($i$ and $j$). The determination of a connection between $i$ and $j$ was done by calculating a partial correlation coefficient adjusted for motion and physiological noises (see [21] for further details).

An unweighted, undirected network was then generated for each subject by applying a threshold to the correlation matrix to yield an adjacency matrix $(A_{ij})$. In order to compare data across people, it is necessary to generate comparable networks. The network was defined so that the relationship between the number of nodes $n$ and the average node degree $K$ is the same across different subjects. In particular, the network was defined so that $S=\log(n)/\log(K)$ is the same across subjects, with $S = 2.8$. This relationship is based on the path length of a random network with $n$ nodes and average degree $K$ [4,5], and can be re-written as $n = K^S$. Our analysis includes ERGM fits to these thresholded whole-brain functional connectivity networks for each subject (an example of which is shown in Figure **1**).

## Model definition

Exponential random graph models have the following form [22]:

$$P_{\boldsymbol{\theta}}(\boldsymbol{Y} = \boldsymbol{y}) = \kappa(\boldsymbol{\theta})^{-1} \exp\{\boldsymbol{\theta}^{\mathrm{T}} \mathbf{g}(\boldsymbol{y})\}. \tag{1}$$

Here $\boldsymbol{Y}$ is an $n \times n$ ($n$ nodes) random symmetric adjacency matrix representing a brain network from a particular class of networks, with $Y_{ij} = 1$ if an edge exists between nodes $i$ and $j$ and $Y_{ij} = 0$ otherwise. Nodes represent locations in the brain (e.g., ROIs) and edges represent functional or structural connections between them. We statistically model the probability mass function (pmf) ($P_{\boldsymbol{\theta}}(\boldsymbol{Y} = \boldsymbol{y})$) of this class of networks as a function of the prespecified network features defined by the $p$-dimensional vector $\mathbf{g}(\boldsymbol{y})$. This vector of explanatory metrics consists of covariates that are functions of the network $\boldsymbol{y}$ and can contain any graph statistic (e.g., number of paths of length two) or node statistic (e.g., brain location of the node). The parameter vector $\boldsymbol{\theta} \in \mathbb{R}^p$, associated with $\mathbf{g}(\boldsymbol{y})$, quantifies the relative significance of the network features in explaining the structure of the network after accounting for the contribution of all other network features in the model and must be estimated. More specifically, $\theta$ indicates the change in the log odds of an edge existing for each unit increase in the corresponding explanatory metric. If the $\theta$ value corresponding to a given metric is large and positive, then that metric plays a considerable role in explaining the network architecture and is more prevalent than in the null model (random network with the probability of an edge existing ($p$) = 0.5). Conversely, if the $\theta$ value is large and negative, then that metric still plays a considerable role in explaining the network architecture but is less prevalent than in the null model. Consequently, inferences can be made about whether certain local features/substructures are observed in the network more than would be expected by chance enabling hypothesis development regarding the biological processes that produce these structural properties. The normalizing constant $\kappa(\boldsymbol{\theta})$ ensures that the probabilities sum to one. This approach allows representing the global network structure by locally specified explanatory metrics, thus providing a means to examine the nature of networks that are likely to emerge from these effects.

The goal in defining $\mathbf{g}(\boldsymbol{y})$ is to identify local metrics that concisely summarize the global (whole-brain) network structure. **Table 1** defines a subset of mathematically compatible explanatory network metrics (for further details see [16,23,24]). Several analogs to these metrics for directed graphs have been detailed by [25]. The GWD, GWESP, and GWDSP statistics discussed in [17] help address degeneracy issues illuminated in [22] and [26]. These issues concern the shape of the estimated pmf (e.g., a pmf in which only a few graphs have nonzero probability) and can lead to lack of model convergence and unreliable results. As noted by [16,27], the most appropriate explanatory metrics vary by network type. Thus, an exploration of which network metrics best characterize brain networks has great appeal. Once the most appropriate statistics have been established, parameter profiles ($\boldsymbol{\theta}$) can be utilized to classify and compare whole-brain networks. These parameter profile comparisons require the use of a uniform set of explanatory metrics for all networks (due to metric interdependencies) and *balanced* networks (same number of nodes for all networks) due to the dependence of the metrics on network size.

It is important to note that ERGMs can be thought of as a way of parameterizing models for networks, and are not a "kind" of network model in the way "model" is traditionally used in the brain network literature. Most other network models, in theory, should have an equivalent ERGM expression (though that specific expression may not be convenient, parsimonious, etc.). For instance, an ERGM with just the Edges metric (**Table 1**) in the formulation (i.e., in $\mathbf{g}(\boldsymbol{y})$) is equivalent to the Erdos-Renyi model. Thus, ERGMs allow parameterizations that subsume most (if not all) other network models.

Fitting of the ERGM in equation 1 is normally done with either Markov chain Monte Carlo maximum likelihood estimation (MCMC MLE) or maximum pseudo-likelihood estimation (MPLE) ([28] contains details). Model fits with MPLE are much simpler computationally than MCMC MLE fits and afford higher convergence rates with large networks. However, properties of the MPLE estimators are not well understood, and the estimates tend to be less accurate than those of MCMC MLE. Here we employ MCMC MLE to fit the model in equation 1 given that there were no convergence issues. See [29] for further details about this estimation approach which can be implemented in the statnet package [23] for the R statistical computing environment.

## Model selection

In order to establish the most appropriate set of explanatory metrics for each subject's brain network and provide a foundation for the development of a "best assessment" ERGM for analyzing complex brain networks, we implemented and assessed three model/metric selection methods. They include a traditional p-value based backward selection approach [30], an information criterion approach (AIC, [31]), and a graphical goodness of fit (GOF) approach [17]. The latter two techniques are used most often for metric selection in ERGMs [16,17]; and, to our knowledge, no detailed comparisons have been performed to determine whether the approaches generally produce the same "best" model. The p-value approach is based on removing metrics that are not statistically significant. Whereas, the AIC approach selects the set of metrics that produce the estimated distribution most likely to have resulted in the observed data with a penalty for additional metrics to ensure parsimony. Alternatively, the graphical GOF method allows subjectively selecting the set of explanatory metrics that produces the model most able to capture and reproduce certain topological properties of the observed network (see Appendix S1 for more details). For each approach ERGMs were fitted to the 90-node unweighted, undirected brain networks of the 10 subjects discussed previously. The potential explanatory metrics ($\mathbf{g}(\boldsymbol{y})$) for each of the 10 networks are listed by category in **Table 2**. The categories were chosen based on properties of brain networks that are regarded as important in the literature [14]. These metrics are analogous to typical brain network metrics (e.g., clustering coefficient ($C$)) but have been developed to be statistically compatible with ERGMs. **Figure 2** illustrates the calculation of the less widely used of these statistics, namely GWESP, GWNSP, and GWDSP, on a six-node example network. The distribution of the unweighted analogues of these metrics (ESP, NSP, and DSP) is given for simplicity. The weighted versions simply sum the values of the distribution giving less weight to those with more shared partners. For this example we note that the network has 1 set of connected nodes with 0 shared partners ($ESP_0$), 5 sets with 1 shared partner ($ESP_1$), 1 set with 2 shared partners ($ESP_2$), and 0 sets with 3 or 4 shared partners ($ESP_3$ and $ESP_4$). Further details on the metrics are provided in **Table 1** and [27]. The $\tau$ parameters associated with GWESP, GWDSP, GWNSP, and GWD were all assumed to be fixed and known (for reasons outlined in [17]) and set to $\tau = 0.75$ based on preliminary analyses as this value generally led to better fitting models according to all selection methods. The three aforementioned model selection approaches are outlined in Appendix S1.

## Results

We implemented the model selection procedures delineated in the previous section and Appendix S1 for each of the 10 subjects using the statnet package [23] for the R statistical computing environment. The resulting models (for each approach) and their corresponding parameter estimates are displayed in **Table 3**. These estimates quantify the relative significance of the given metric in explaining the overall network structure; and, more specifically, they specify how much the log odds of an edge existing increases for each unit increase in the corresponding metric. For example, the final graphical GOF model for subject 10 shows that GWESP is the most important metric (other than the number of edges) in describing the structure of the subject's network given the larger absolute value of the parameter estimate. Additionally, the positiveness of the estimate associated with GWESP indicates that an edge that closes a triangle is more likely to exist than it would by chance (i.e., the network has more clustering than a random network where the probability of an edge is $p = 0.5$) for the family of networks represented by subject 10's fitted model. As evidenced by the results in **Table 3**, the three model selection methods can lead to very different "best" models. The disparate final model GOF plots that can result from the three different model selection approaches are exhibited in **Figures 3** and **4** (for subjects 2 and 8). Again, our aim here is not to judge the three selection methods, but to highlight the fact that they can lead to disparate final models/sets of features. These model selection approaches have been used seemingly arbitrarily in the literature; and, to our knowledge, no detailed comparisons have been performed to determine whether the approaches generally produce the same "best" model/set of features. For our purposes we recommend the graphical GOF approach as the standard and will use it in future analyses given that our main scientific interest lies in being able to capture and reproduce the structure of the fitted brain networks. With the exception of subject 8, the graphical GOF approach produces reasonably good fits for all subjects. The remaining best graphical selection model GOF plots are shown in **Figures 5-12**.

Despite the obvious importance of Edges (as evidenced by the absolute values of its parameter estimates in **Table 3**) in the models, the overlap between the simulated and observed networks in the GOF plots is not merely an effect of pure connectivity, but also an effect of network organization. As mentioned in the Materials and Methods section, an ERGM with just an Edges metric is equivalent to the Erdos-Renyi random graph. Thus, due to the small worldness of brain networks, models of this type will not capture the tight local clustering/regional specificity (among other properties) present in these networks [5]. **Figure 13** illustrates this point by exhibiting the disparate GOF plots for an Edges only model and the final graphical selection model for subject 10. Clearly the Edges only model is unable to capture the regional specificity (edge-wise shared partners distribution) and global processing (minimum geodesic distance distribution) properties of brain networks which are well embodied by the final graphical selection model.

The bolded explanatory metrics in **Table 3** are those contained in at least half ( $\geq 5$ ) of the "best" subject network models based on the graphical GOF approach. Examining the uniformity of the selected explanatory metrics across subjects in this way is needed for the development of a "best assessment" ERGM for the reasons detailed in the Introduction and Materials and Methods sections. Examination of these metrics leads to an overall ERGM for whole-brain networks that requires a Connectedness metric (Edges), a Local Efficiency metric (GWESP), and a Global Efficiency metric (GWNSP). That is,

$$P_{\boldsymbol{\theta}}(\boldsymbol{Y} = \boldsymbol{y}) = \kappa(\boldsymbol{\theta})^{-1}\exp\{\theta_1\text{Edges} + \theta_2\text{GWESP} + \theta_3\text{GWNSP}\}. \qquad (2)$$

These three metrics having the most influential impact on overall functional brain network organization in these subjects seems consistent with our biological understanding of the brain. The number of functional connections present (Edges) is clearly instrumental in information transfer while also playing a role in brain network organization [6]. Clustering (GWESP) is another critical feature of brain network architecture that allows the efficient local processing of information. The consistently negative $\theta_3$ values associated with GWNSP indicate that if two brain areas are not functionally connected, they are less likely to have shared connections with other regions than they would by chance. That is, two regions are less likely than by chance to have a 2-path as the shortest path between them. Speculatively, this may result from the brain having direct connections when necessary, but allowing for slightly longer global connections (3-paths, etc.) to maintain efficiency otherwise. Additionally, the synergistic combination of these metrics engenders networks that well capture the geodesic (global efficiency), shared partner (local efficiency), degree, and triad census (motifs) distributions of brain networks as evidenced by several of the GOF plots in Figures **3-13**.

Group-based network comparisons can potentially be performed by comparing the mean of the estimated $\theta_1$, $\theta_2$, and $\theta_3$ values among groups via hypothesis testing or classification techniques. It is important to note that if one were to just compare the mean of the estimated $\theta_1$ (Edges) values among groups, for instance, potential confounding from the GWESP and GWNSP would be inherently accounted for given that the estimates account for all other metrics in the model. In the hypothesis testing framework one can exploit the fact that the $\theta$'s are approximate MLEs and thus asymptotically have a Gaussian distribution. Approximate T-tests and/or F-tests can then be employed. Investigating the individual differences in final models among subjects is also important. Although parameter values cannot be directly compared when different models are fitted, the disparate fits themselves may elucidate biologically interesting differences among groups or individual subjects.

Here we implement our best assessment ERGM from equation 2 to illustrate its utility for comparing groups of networks. The subjects were split into a younger (aged 20-26) and slightly older (aged 29-35) group (5 subjects each) in order to assess if there were any discernible differences between their brain networks. Other studies have shown that older adults tend to have less clustering and slightly more connections than their younger counterparts [32,33]. However, direct comparisons have not been done on groups of subjects this close in age to establish whether these changes tend to commence immediately or take effect at older ages. Moreover, these studies did not consider the potential confounding effects of other network metrics when assessing these differences. As evidenced by the results of our analysis exhibited in **Table 4**, the two groups differ significantly in $\theta_3$ (the GWNSP parameter) with the younger group having a more negative value. That is, if two nodes are not functionally connected, they are more likely to have shared connections with other nodes in the brain networks of the older group. Biologically, this could be the result of the older brain maintaining two-path connections between brain areas that have lost their direct connections; however, this interpretation is purely speculative at this point. Interestingly, there is not a statistically significant difference between the groups for the Edges or GWESP parameter, with the trend being for the older subjects' networks to have more connections and clustering. These findings run counter to those in the literature and may stem from the fact that our analysis accounts for some of the confounding that arises from network metric dependencies [6,7]. These disparate findings could also just be a result of the closeness in age of the two groups or random variability given our small sample size. As noted by [7], larger and methodologically more comparable future investigations are needed to resolve many of the contradictory findings in functional connectivity studies.

In addition to model representation and comparison, ERGMs also provide a statistically sound method for simulating complex brain networks as is done for the GOF plots. To illustrate their utility in this context we simulated 100 networks based on the fitted ERGM of subject 10. We then calculated several descriptive metrics commonly used in the neuroimaging literature for the observed and simulated networks to assess the utility of the simulated networks within the neuroscientific context. **Table 5** displays the results of these computations for Clustering coefficient ($C$), Characteristic path length ($L$), Local Efficiency ($Eloc$), Global Efficiency ($Eglob$), and Mean Nodal Degree ($K$) (see [4,34] for details on these metrics). As evidenced by the results in this table, the simulated networks are very similar to the observed network. Hence ERGMs render an approach to simulating scientifically meaningful brain networks.

## Discussion

Our analyses in the previous section illustrate the utility of ERGMs for modeling, analyzing, and simulating complex whole-brain networks. We have also provided a foundation for the development of a best assessment ERGM for the classification and comparison of brain networks via the evaluation of three model/feature selection approaches. The graphical GOF approach serves as the best method given the scientific interest in being able to capture and reproduce the structure of the fitted networks. The greatest appeal of modeling brain networks with ERGMs lies in their ability to efficiently represent this complex network data and allow examining the way in which a network's global structure and function depend on local structural components.

There are a myriad of ways in which ERGMs can potentially be useful for brain network researchers. As previously discussed and demonstrated, groups of networks can be statistically compared and classified (by disease status, age, task, etc.) based on several network features simultaneously. The models also provide a way of exploring which local features of brain networks are most important in explaining their global architecture. As noted by many authors [35-38], an analysis approach that can capture the network characteristics from a group of subjects' brain networks is needed. ERGMs provide a potential solution since one could average the parameter profiles, $\boldsymbol{\theta}$, of a group and then simulate "representative" networks based on this averaged profile. Preliminary work has shown this approach to be quite effective. These representative networks can serve as null networks against which other networks and network models can be compared, as visualization tools, and as a means for characterizing properties of network metrics in a group (e.g., community structure). ERGMs, in general, will also serve to both accommodate the ever increasing complexity of whole-brain analyses and inform future statistical models for whole-brain research.

A computational limitation of note for brain network researchers is that MCMC MLE fits of ERGMs can be computationally intensive and may fail to converge with more spatially resolved networks than the 90 ROI ones used here. This fitting algorithm has been shown to handle networks of several thousand nodes [17]; however, its effectiveness is more dependent on the number and topological structure of the edges than the node count [23]. Future work will examine the scalability of ERGMs fitted with MCMC MLE in the context of brain networks. As convergence issues arise with more finely parcellated networks, MPLE fits may serve as an appropriate alternative [16].

Another potential issue of note is that the original data's variability may affect the resulting ERGM fits. A given subject may exhibit variability of the connections in their brain networks at different times of day due to experimental or physiological reasons. Assessment of the robustness of ERGM fits to this within-subject variability is important and will be the focus of future investigations.

In addition to the utility of ERGMs in the research context, the potential implications of their use in the clinical context are profound as they can aid in elucidating system level functional features/neurological processes (represented by the explanatory network metrics) that play a role in various cognitive disorders. For instance, several authors have shown that schizophrenics have less local efficiency in their brain networks [7,35,36] (which would correspond to a smaller parameter estimate for GWESP in the ERGM framework) than control subjects. ERGMs enable empirically examining how this difference in efficiency affects global brain structure and comparing these emergent whole-brain brain networks between schizophrenics and controls. For example, one could simulate networks based on model fits to schizophrenics and controls to see how this difference affects the variability of the resulting networks. This comparison may give us insight into the neurological mechanisms that lead to schizophrenia (e.g., lack of local neuronal communication leads to less stability in global structure for schizophrenics).

Aside from the aforementioned clinical and biological work that can be done with the models, there are also many possible directions for future methodological research involving the analysis of complex brain networks with ERGMs. Approximating the small-sample distribution of $\boldsymbol{\theta}$ may prove useful for hypothesis testing frameworks in which appealing to asymptotic normality may not be appropriate. Developing methods for quantifying GOF plots to remove subjectivity and allow for analytical comparisons of the graph will be valuable. The approach should allow some flexibility in determining how to weight the four comparison statistics with respect to their relative importance to the scientific context. Developing novel explanatory network metrics rooted in both the biology of the brain and the mathematics of ERGMs will engender better best assessment models for network comparison. A corresponding hybrid model selection approach where models are penalized for using many covariates and the GOF plots are assessed will prove useful in maintaining parsimony as the number of relevant explanatory metrics increases. The extension of ERGMs to directed and/or weighted brain networks will prove beneficial as construction of these network types gains feasibility.

## Acknowledgements

We thank the editor and referees for their comments that considerably improved the paper. An earlier version of this manuscript can be found at arxiv.org (Simpson et al., arXiv:1007.3230v1 [stat.AP]).

**Table 1.** Subset of explanatory network metrics

| Metric | Description |
|---|---|
| Edges | Number of edges in network |
| Two-Path | Number of paths of length 2 in the network |
| k-Cycle | Number of k-cycles in network |
| k-Degree | Number of nodes with degree k |
| Geometrically weighted degree (GWD) | Weighted sum of the counts of each degree $(i)$ weighted by the geometric sequence $(1 - \exp\{-\tau\})^i$, where $\tau$ is a decay parameter |
| Geometrically weighted edge-wise shared partner (GWESP) | Weighted sum of the number of connected nodes having exactly $i$ shared partners weighted by the geometric sequence $(1 - \exp\{-\tau\})^i$, where $\tau$ is a decay parameter |
| Geometrically weighted non-edge-wise shared partner(GWNSP) | Weighted sum of the number of non-connected nodes having exactly $i$ shared partners weighted by the geometric sequence $(1 - \exp\{-\tau\})^i$, where $\tau$ is a decay parameter |
| Geometrically weighted dyad-wise shared partner (GWDSP) | Weighted sum of the number of dyads[a] having exactly $i$ shared partners weighted by the geometric sequence $(1 - \exp\{-\tau\})^i$, where $\tau$ is a decay parameter |
| Nodematch | Number of edges $(i, j)$ for which nodal attribute $i$ equals nodal attribute $j$ (e.g., brain location of node $i$ = brain location of node $j$) |

[a]node pair with or without edge

**Table 2.** Explanatory network metrics by category

| Category | Metric(s) |
|---|---|
| 1) Connectedness | Edges, Two-Path |
| 2) Local Clustering/Efficiency | GWESP, GWDSP |
| 3) Global Efficiency | GWNSP[a] |
| 4) Degree Distribution | GWD |
| 5) Location (in the brain) | Nodematch |

NOTE: See Table 1 for more details on the metrics.
[a]Not inherently global, but helps produce models that accurately capture the global efficiency of our networks.

**Table 3.** Final model estimates by model selection approach for each subject

| Subject | Approach | Final Model[a] | | | | | | |
|---|---|---|---|---|---|---|---|---|
| | | **Edges** | Two-Path | **GWESP** | GWDSP | **GWNSP** | GWD | Nodematch |
| 2 | p-value | −2.29 | − | − | 0.90 | −1.12 | −1.53 | 1.11 |
| | AIC | −2.29 | − | − | 0.90 | −1.12 | −1.53 | 1.11 |
| | Graphical | −2.85 | − | 1.00 | − | −0.28 | − | 0.93 |
| 3 | p-value | − | 0.12 | − | − | −0.64 | −1.83 | 1.53 |
| | AIC | − | − | 0.30 | − | −0.48 | −2.14 | 1.50 |
| | Graphical | −2.09 | − | − | 0.72 | −1.08 | − | 1.59 |
| 5 | p-value | −3.07 | − | − | 1.02 | −1.30 | − | 1.25 |
| | AIC | −3.09 | − | 0.99 | − | −0.28 | − | 1.24 |
| | Graphical | −3.09 | − | 0.99 | − | −0.28 | − | 1.24 |
| 8 | p-value | − | −0.03 | − | 0.45 | −1.22 | − | 0.80 |
| | AIC | −3.48 | − | 1.27 | − | −0.28 | − | 1.23 |
| | Graphical | − | 0.05 | − | − | −0.53 | −2.24 | − |
| 9 | p-value | −3.42 | − | − | 1.04 | −1.20 | − | 1.19 |
| | AIC | −2.95 | − | 0.92 | − | −0.21 | −0.67 | 1.21 |
| | Graphical | −3.18 | − | − | 1.06 | −1.23 | − | − |
| 10 | p-value | −4.98 | − | − | 1.52 | −1.62 | 1.19 | 1.41 |
| | AIC | −4.11 | − | − | 1.34 | −1.49 | − | 1.40 |
| | Graphical | −4.48 | − | 1.51 | − | −0.15 | 1.12 | − |
| 12 | p-value | −2.66 | − | − | 0.85 | −1.11 | − | 1.32 |
| | AIC | − | 0.00 | − | 0.33 | −0.80 | −2.61 | 1.22 |
| | Graphical | −2.40 | − | − | 0.87 | −1.14 | − | − |
| 13 | p-value | −2.95 | − | − | 1.08 | −1.37 | − | 1.12 |
| | AIC | −2.96 | − | 1.05 | − | −0.30 | − | 1.11 |
| | Graphical | −2.79 | − | 1.04 | − | −0.30 | − | − |
| 16 | p-value | − | −0.04 | − | 0.41 | −1.07 | − | 0.82 |
| | AIC | −2.29 | − | − | 0.87 | −1.19 | −0.56 | 1.32 |
| | Graphical | −2.25 | − | 0.81 | − | −0.33 | −0.40 | − |
| 21 | p-value | −2.57 | − | − | 0.93 | −1.36 | − | 1.93 |
| | AIC | −1.34 | − | 0.66 | − | −0.40 | −1.56 | 2.05 |
| | Graphical | −2.58 | − | 0.90 | − | −0.35 | − | 1.93 |

[a]Bolded metrics are those contained in at least half of the "best" subject network models based on the graphical GOF approach

**Table 4.** Results of ERGM parameter estimate comparisons between younger and older subjects

| | Younger | | Older | | |
|---|---|---|---|---|---|
| | Mean | SE | Mean | SE | P-value |
| $\theta_1$ (Edges) | $-2.45$ | $3.95 \times 10^{-1}$ | $-3.09$ | $3.47 \times 10^{-1}$ | 0.2626 |
| $\theta_2$ (GWESP) | 0.89 | $1.81 \times 10^{-1}$ | 1.14 | $1.53 \times 10^{-1}$ | 0.3339 |
| $\theta_3$ (GWNSP) | $-0.32$ | $6.62 \times 10^{-3}$ | $-0.24$ | $4.79 \times 10^{-3}$ | $< 0.0001$ |

**Table 5.** Network metrics of observed and simulated networks from subject 10

| Metric | Observed Value | Simulated Networks Mean (SE) |
|---|---|---|
| Clustering coefficient ($C$) | 0.447 | 0.468 (0.004) |
| Characteristic path length ($L$) | 3.520 | 3.475 (0.033) |
| Local Efficiency ($Eloc$) | 0.555 | 0.576 (0.004) |
| Global Efficiency ($Eglob$) | 0.284 | 0.290 (0.003) |
| Mean Nodal Degree ($K$) | 5.066 | 4.939 (0.042) |

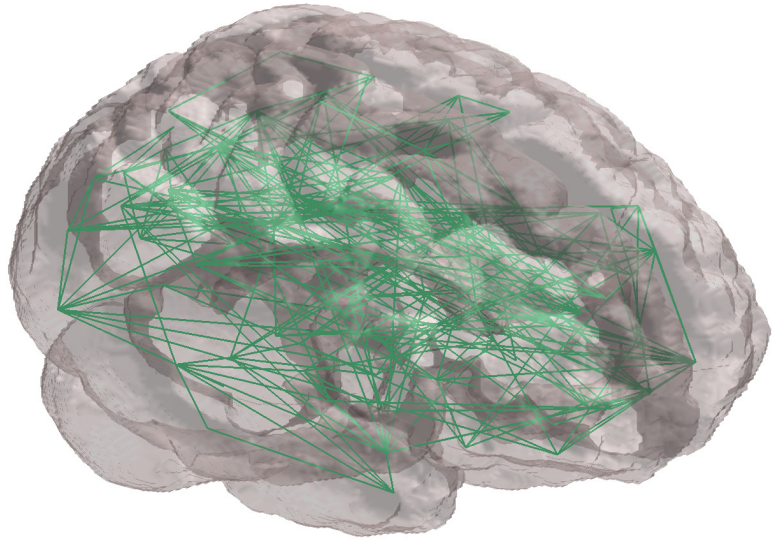

**Figure 1. Network of subject 10 in brain space.**

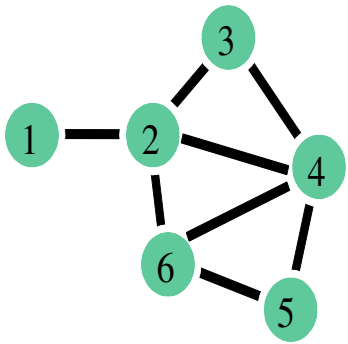

**Figure 2. Six-node example network.** The edgewise, nonedgewise, and dyadwise shared partner distributions are $(\text{ESP}_0, \dots, \text{ESP}_4) = (1, 5, 1, 0, 0)$, $(\text{NSP}_0, \dots, \text{NSP}_4) = (1, 4, 3, 0, 0)$, and $(\text{DSP}_0, \dots, \text{DSP}_4) = (2, 9, 4, 0, 0)$ respectively.

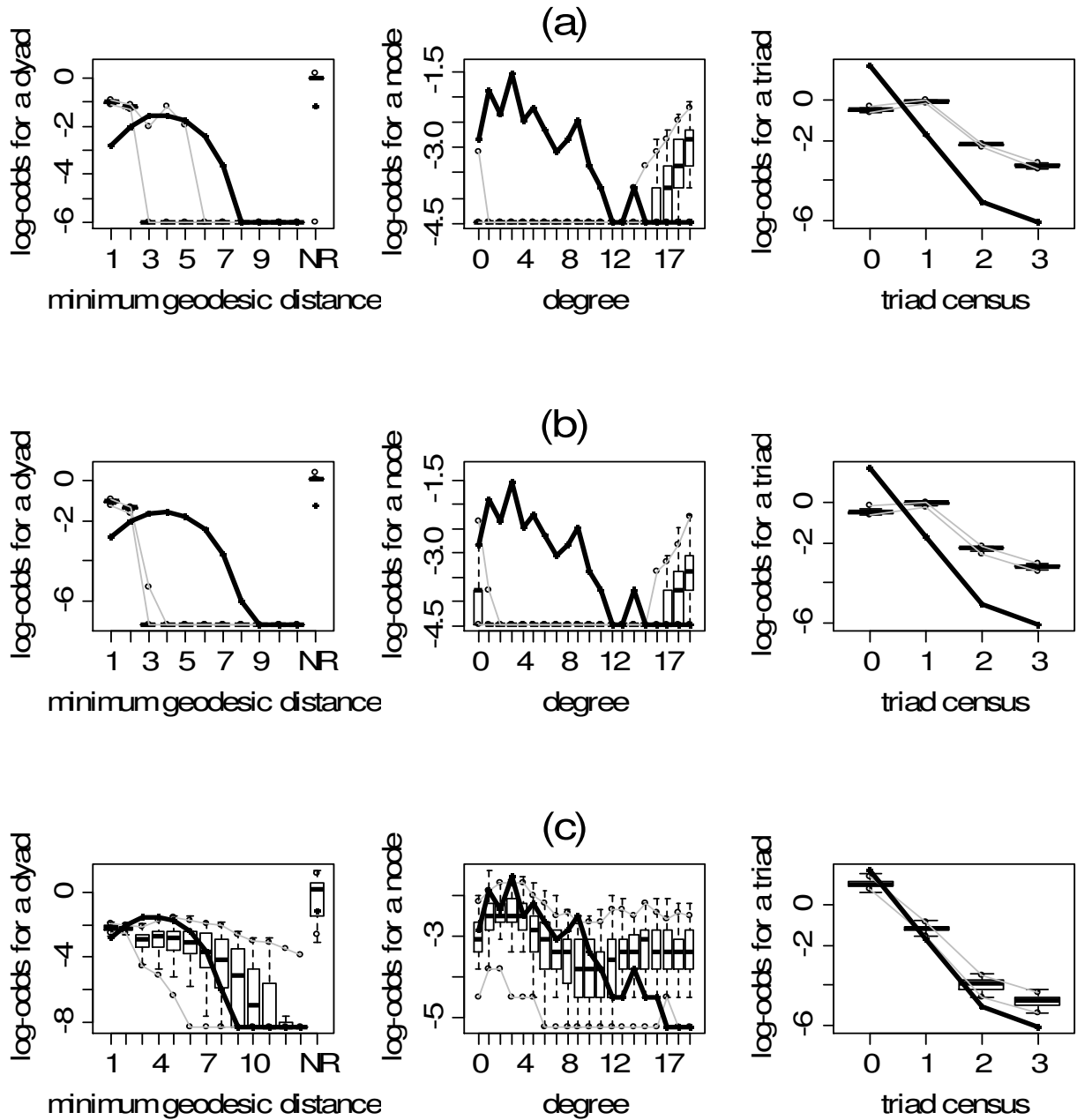

**Figure 3. Goodness-of-fit plots for the final models for subject 2.** The vertical axis is the logit of relative frequency, the solid lines represent the statistics of the observed network, and the boxplots represent the distributions of the 100 simulated networks. (a) P-value model. (b) AIC model. (c) Graphical model.

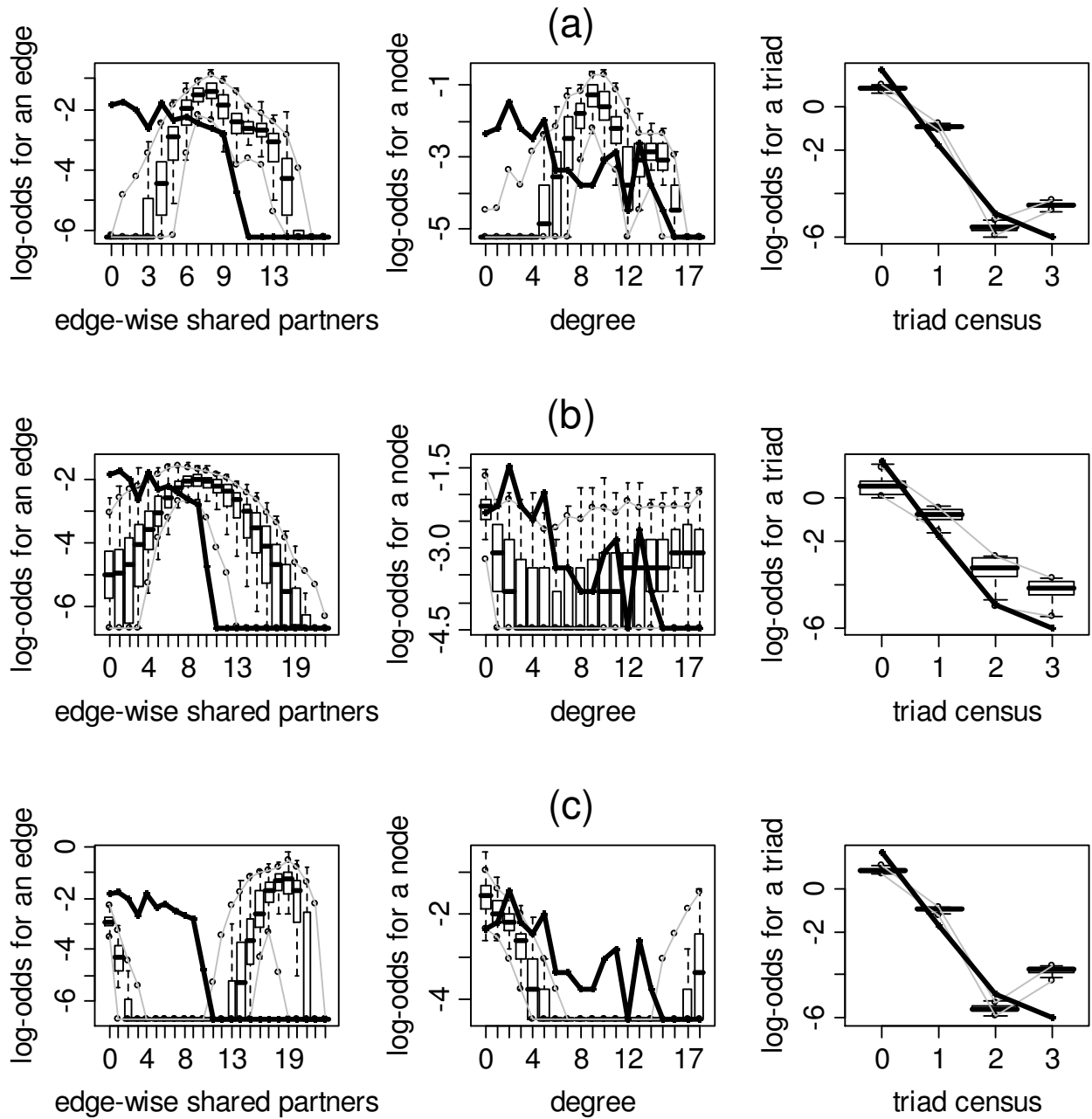

**Figure 4. Goodness-of-fit plots for the final models for subject 8.** The vertical axis is the logit of relative frequency, the solid lines represent the statistics of the observed network, and the boxplots represent the distributions of the 100 simulated networks. (a) P-value model. (b) AIC model. (c) Graphical model.

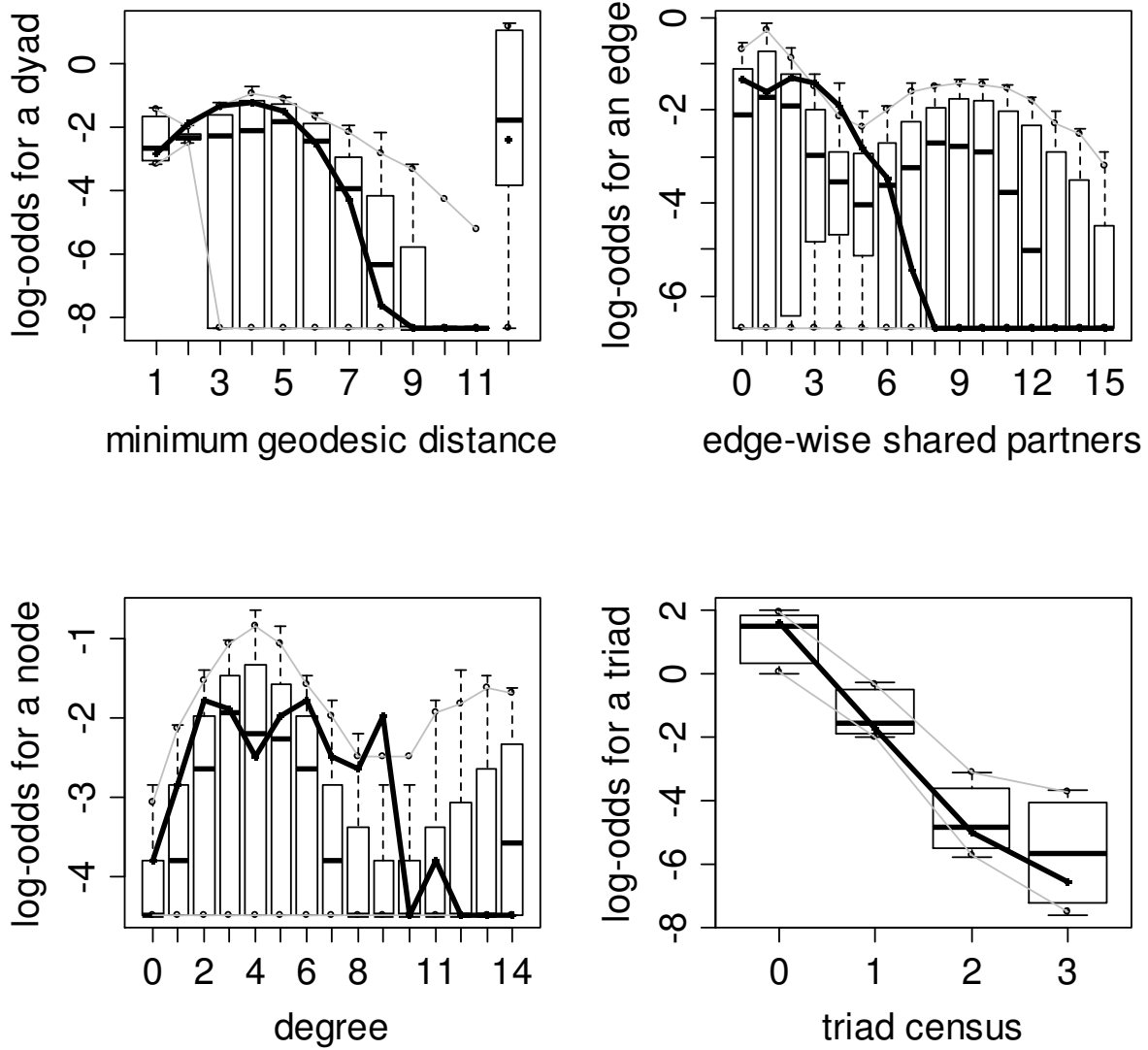

**Figure 5. Goodness-of-fit plots for the final graphical selection model for subject 3.** The vertical axis is the logit of relative frequency, the solid lines represent the statistics of the observed network, and the boxplots represent the distributions of the 100 simulated networks.

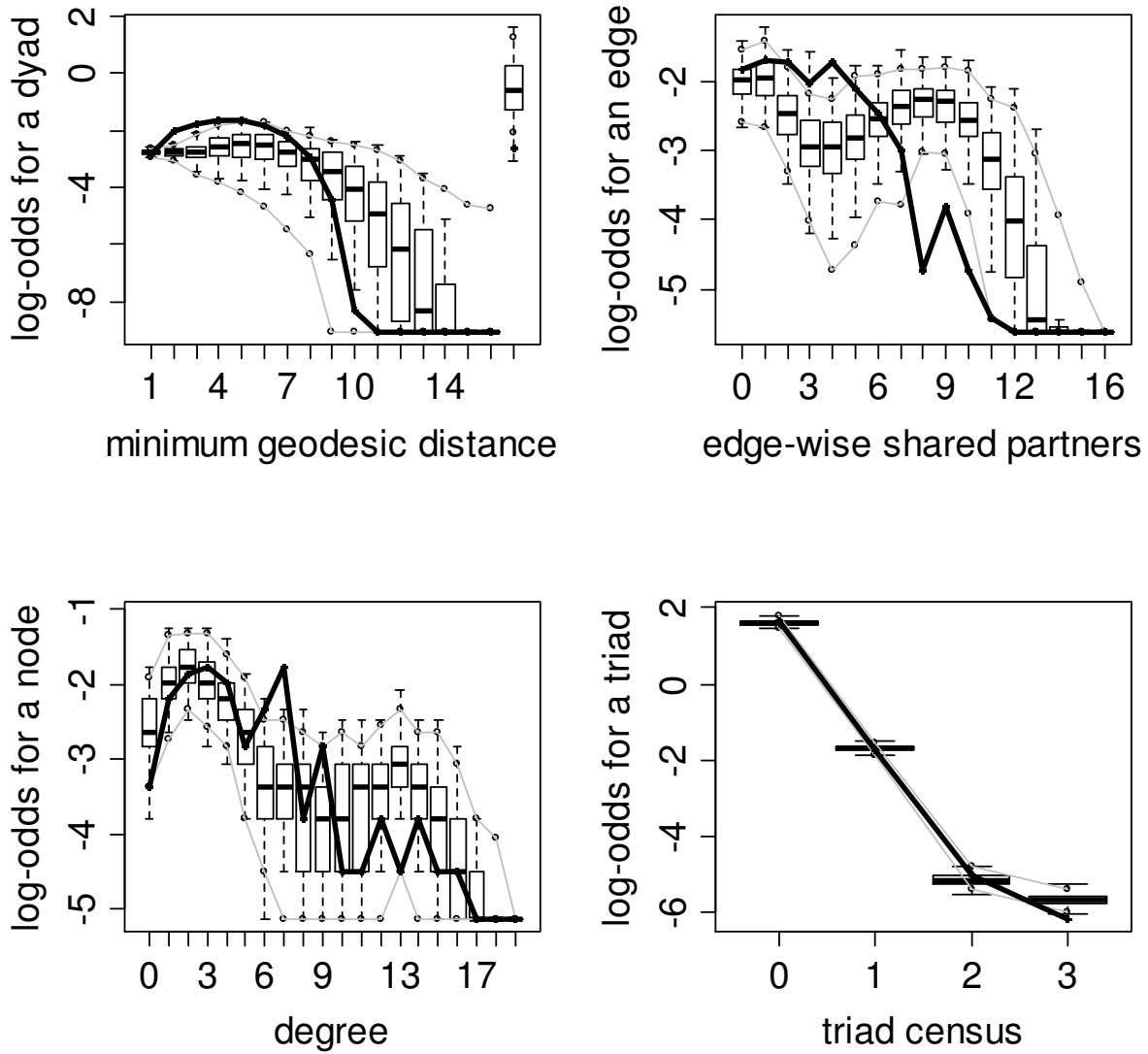

**Figure 6. Goodness-of-fit plots for the final graphical selection model for subject 5.** The vertical axis is the logit of relative frequency, the solid lines represent the statistics of the observed network, and the boxplots represent the distributions of the 100 simulated networks.

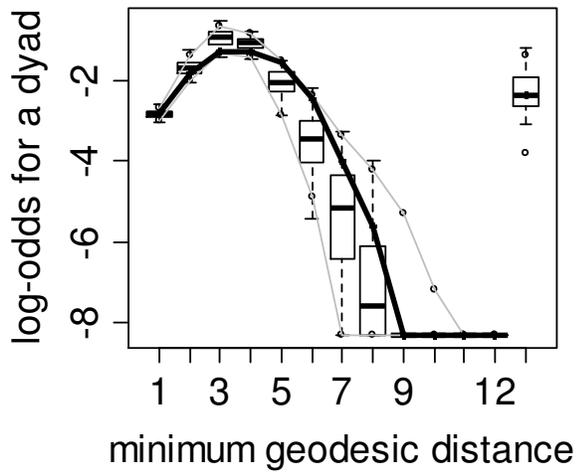
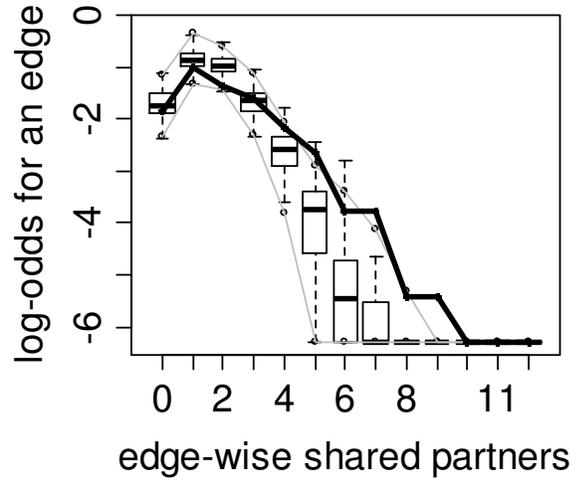
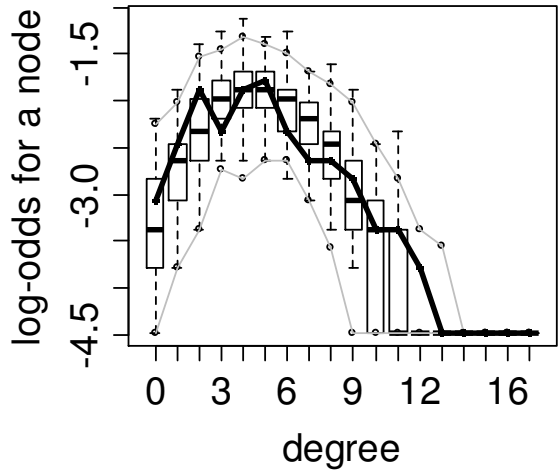
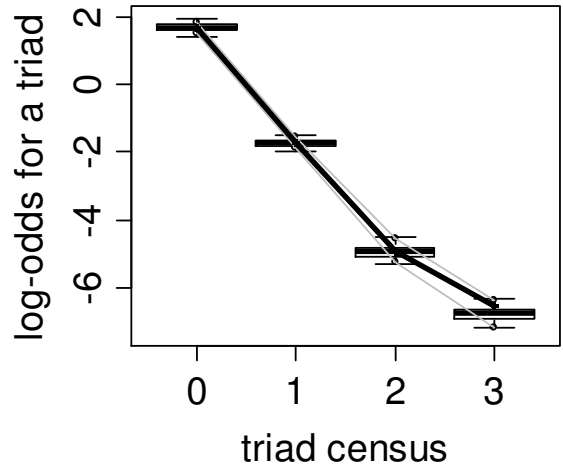

**Figure 7. Goodness-of-fit plots for the final graphical selection model for subject 9.** The vertical axis is the logit of relative frequency, the solid lines represent the statistics of the observed network, and the boxplots represent the distributions of the 100 simulated networks.

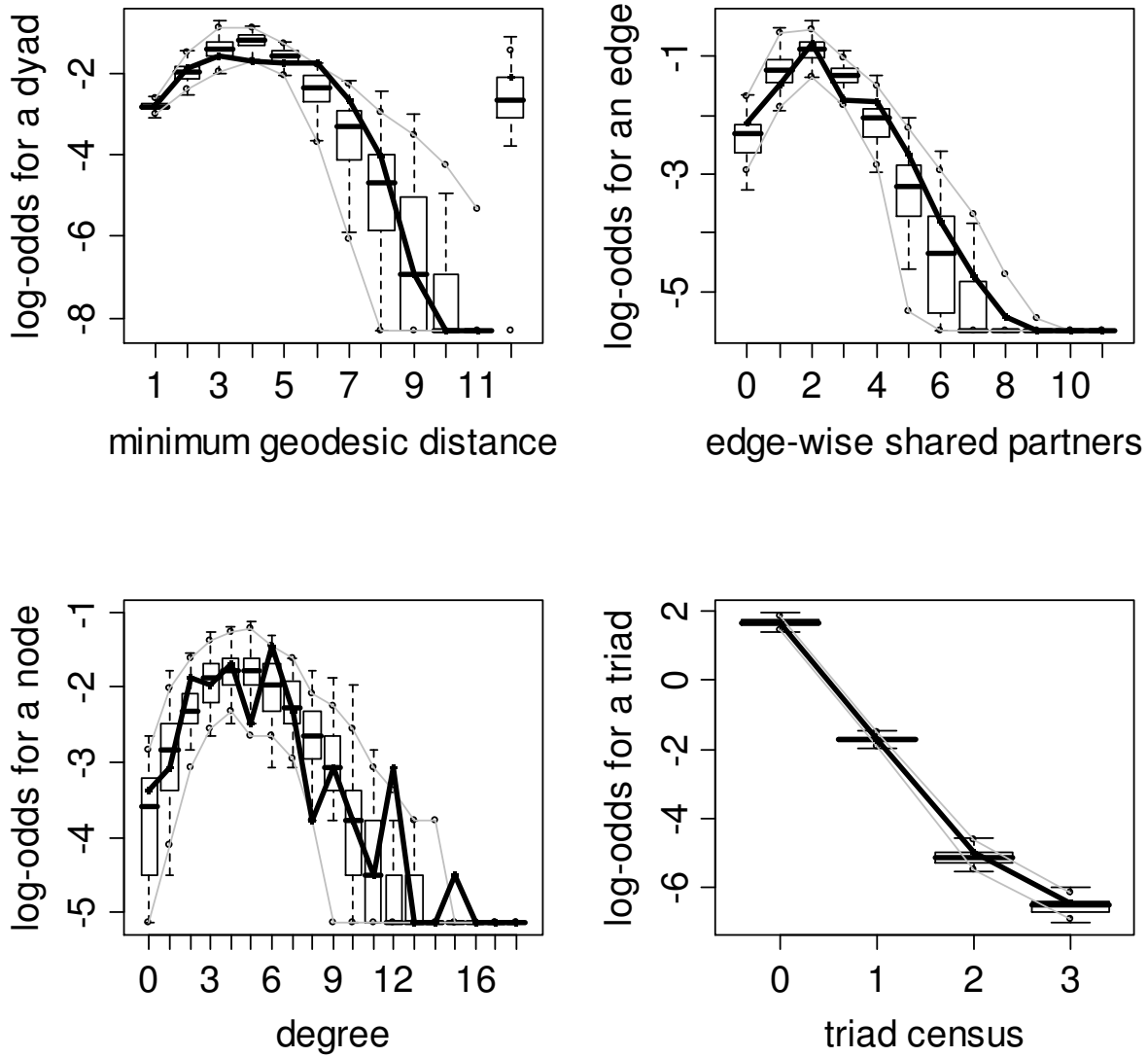

**Figure 8. Goodness-of-fit plots for the final graphical selection model for subject 10.** The vertical axis is the logit of relative frequency, the solid lines represent the statistics of the observed network, and the boxplots represent the distributions of the 100 simulated networks. This exemplifies a good fitting model.

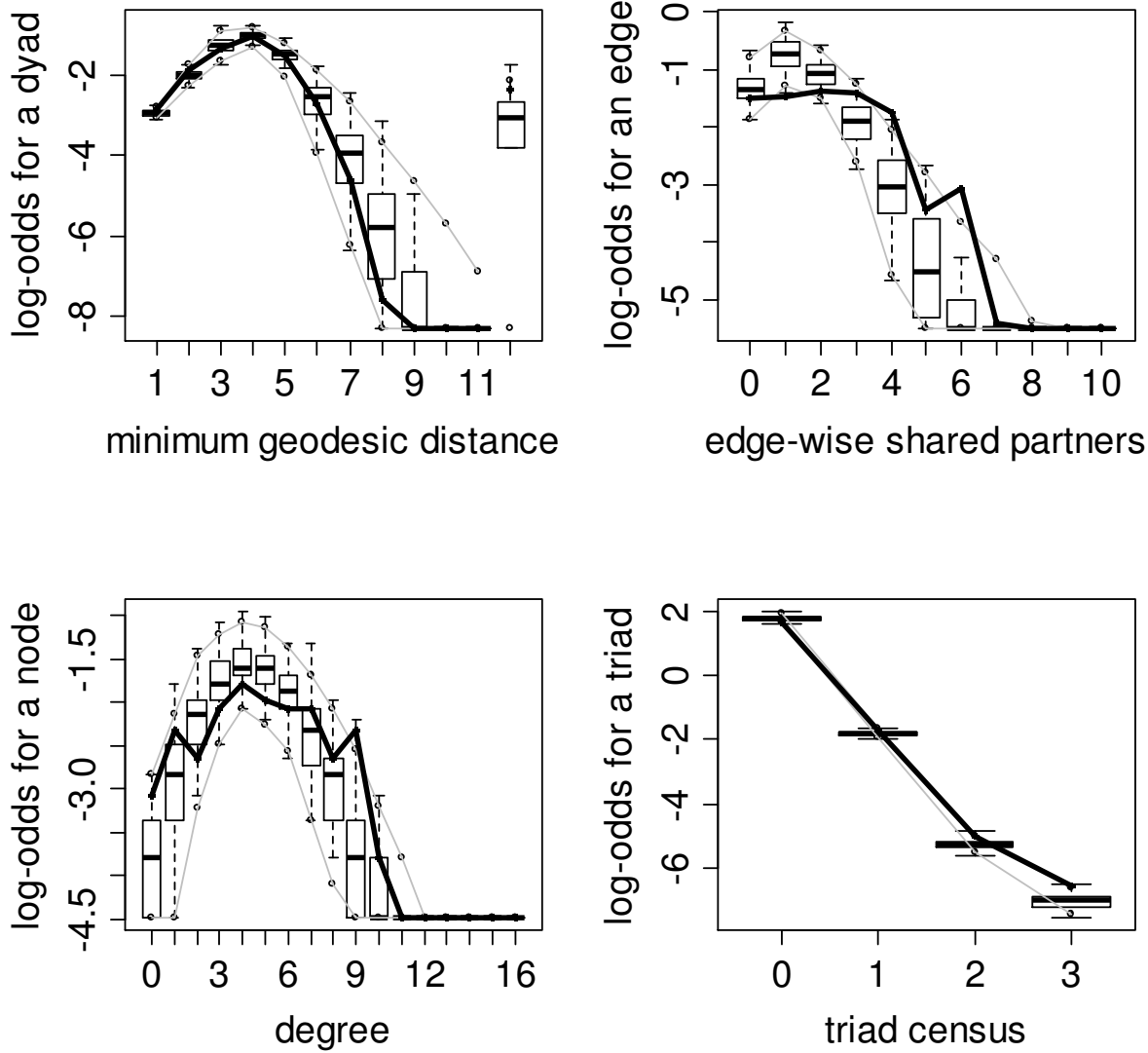

**Figure 9. Goodness-of-fit plots for the final graphical selection model for subject 12.** The vertical axis is the logit of relative frequency, the solid lines represent the statistics of the observed network, and the boxplots represent the distributions of the 100 simulated networks.

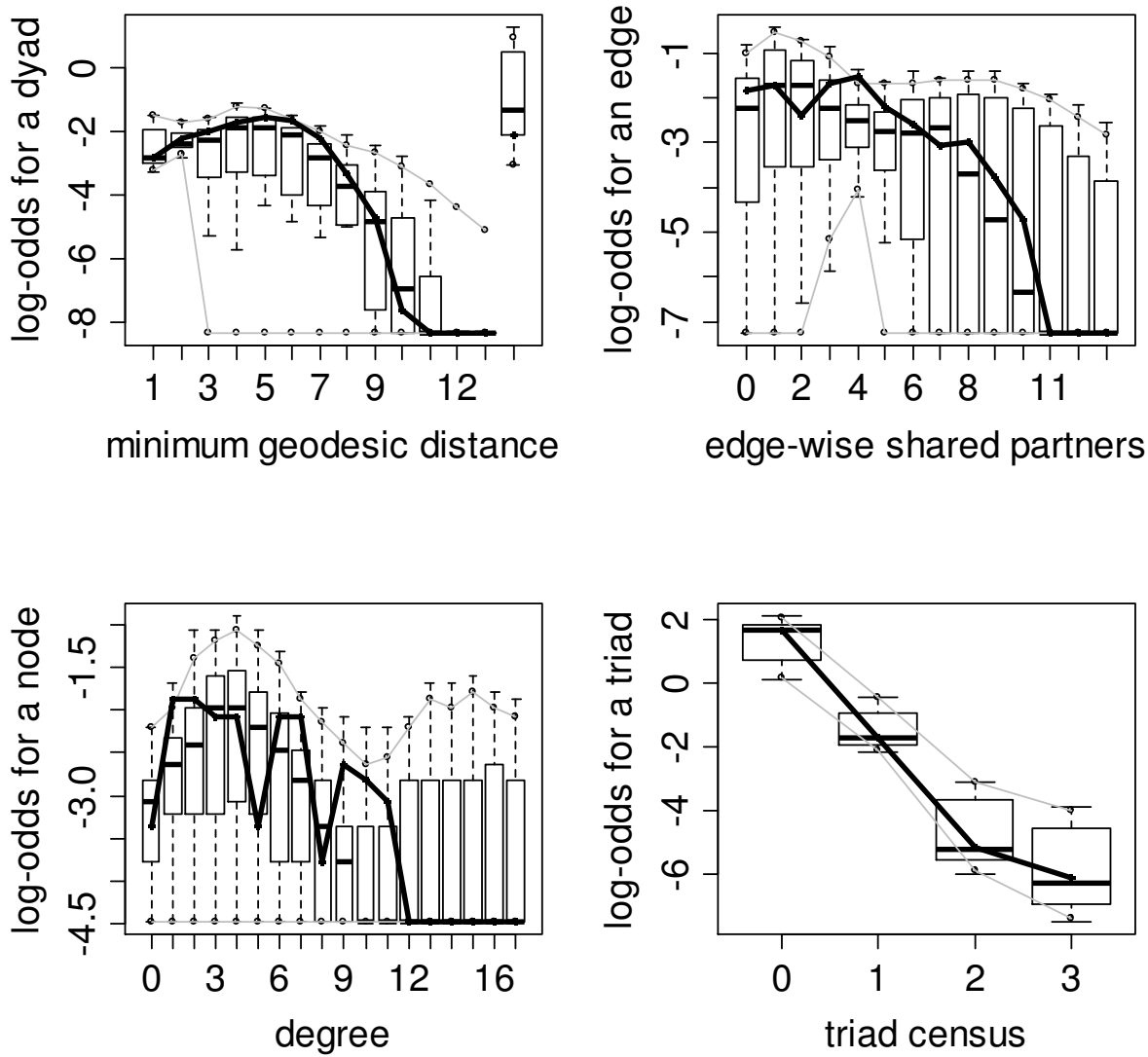

**Figure 10. Goodness-of-fit plots for the final graphical selection model for subject 13.** The vertical axis is the logit of relative frequency, the solid lines represent the statistics of the observed network, and the boxplots represent the distributions of the 100 simulated networks.

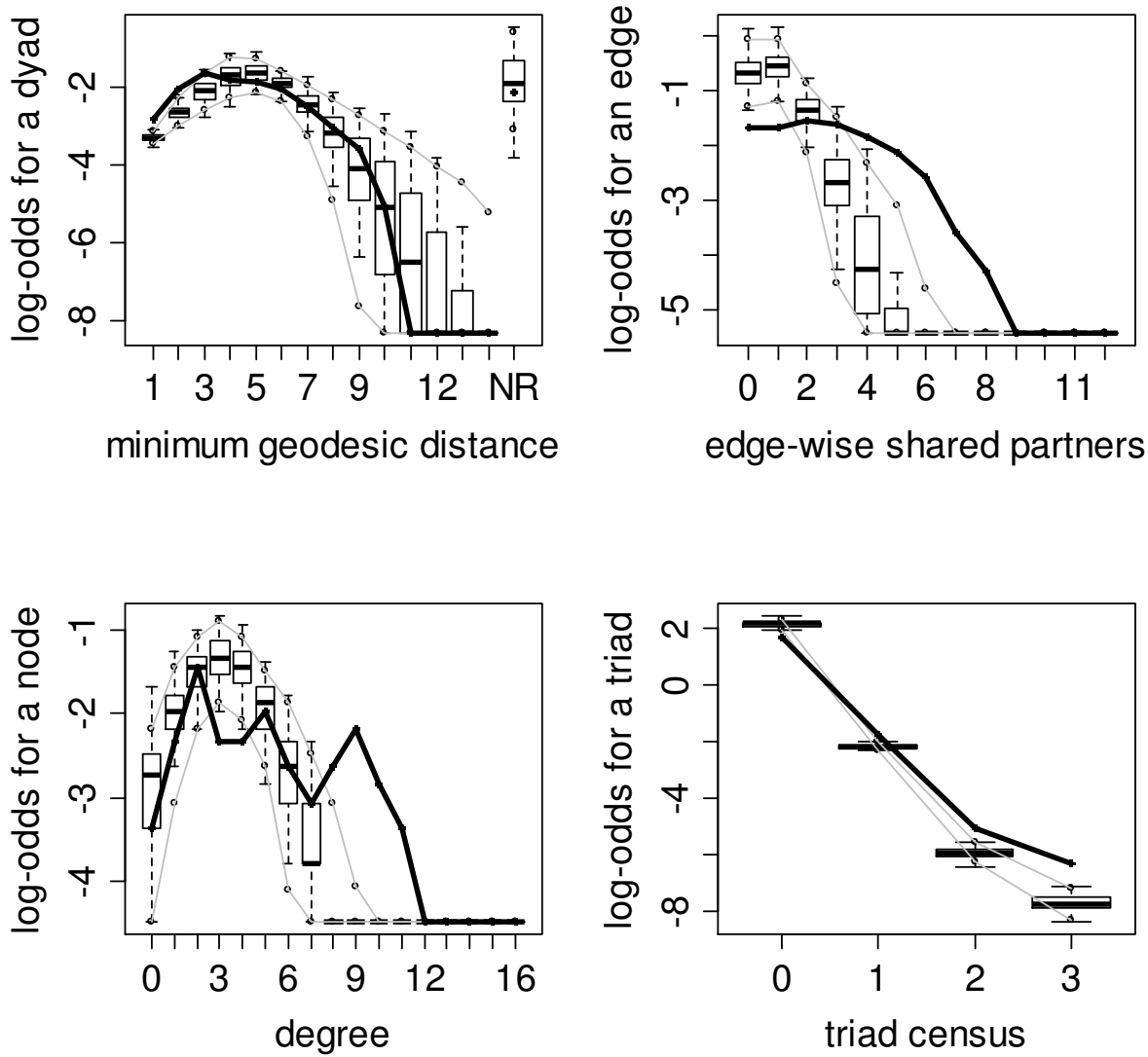

**Figure 11. Goodness-of-fit plots for the final graphical selection model for subject 16.** The vertical axis is the logit of relative frequency, the solid lines represent the statistics of the observed network, and the boxplots represent the distributions of the 100 simulated networks.

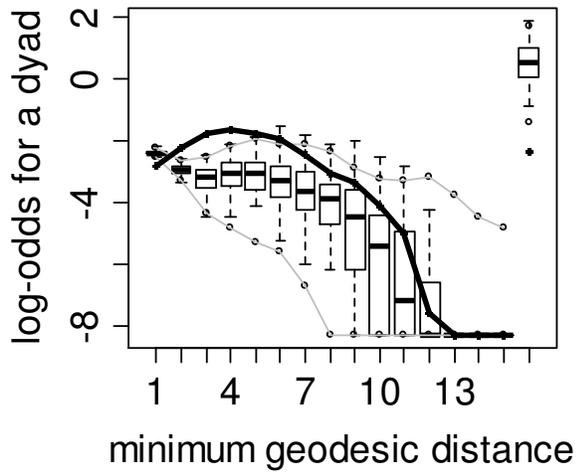
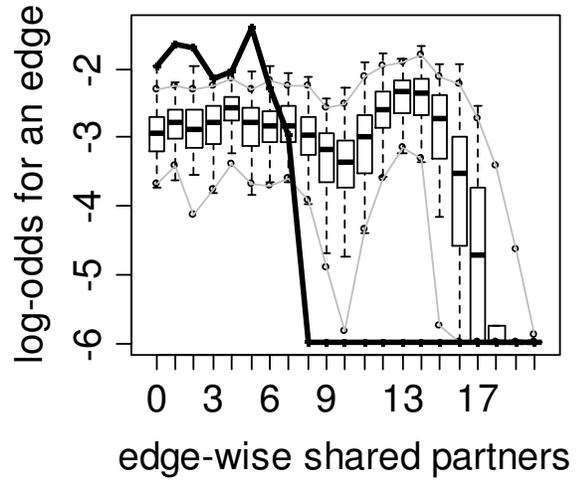
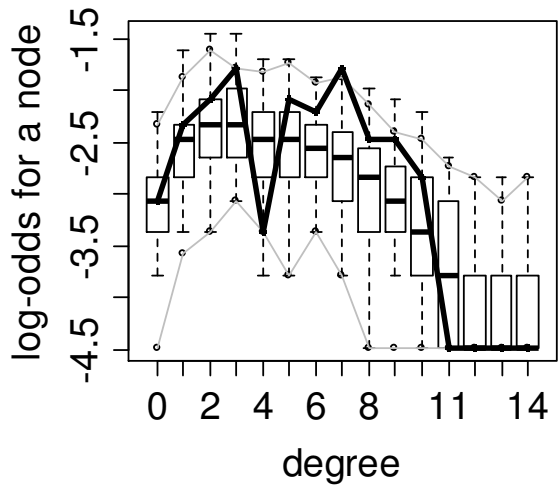
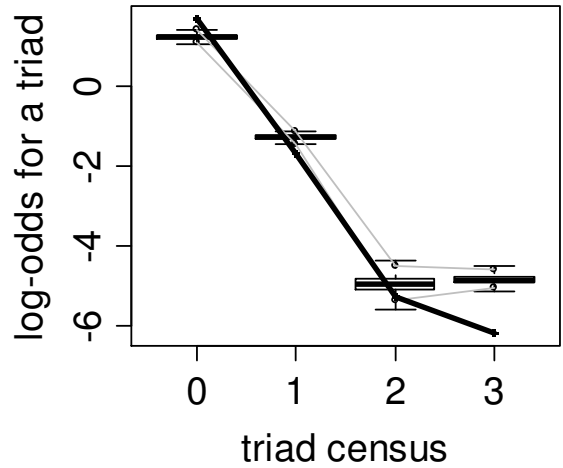

**Figure 12. Goodness-of-fit plots for the final graphical selection model for subject 21.** The vertical axis is the logit of relative frequency, the solid lines represent the statistics of the observed network, and the boxplots represent the distributions of the 100 simulated networks.

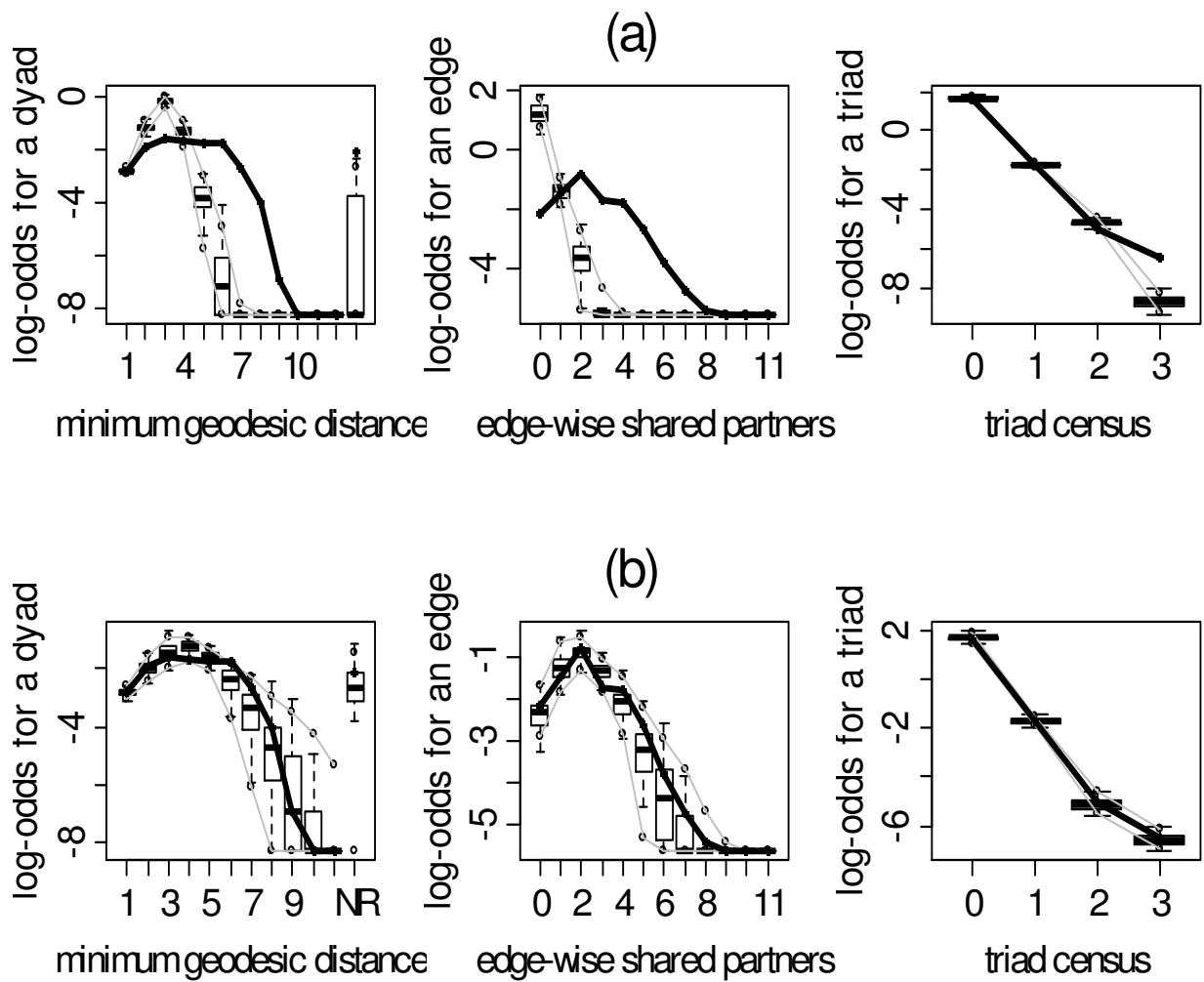

**Figure 13. Goodness-of-fit plots for the Edges only and final graphical selection models for subject 10.** The vertical axis is the logit of relative frequency, the solid lines represent the statistics of the observed network, and the boxplots represent the distributions of the 100 simulated networks. (a) Edges only model. (b) Final graphical model.